\documentstyle[12pt]{article}

\title{Classical and Quantum Chaos\\from Continuous Quantum
Measurements\\
{\normalsize (to be published in ``Solitons, Chaos \& Fractals'')}}

\author{Michael Mensky\thanks{Permanent Address: P.N.Lebedev
Institute, 117924 Moscow, Russia}\\
{\normalsize Fakult\"at f\"ur Physik der Universit\"at Konstanz }\\
{\normalsize D - 78434 Konstanz}\\
{\normalsize Germany}}
\date{November 20, 1993}

\newcommand{\be}{\begin{equation}}
\newcommand{\ee}{\end{equation}}
\newcommand{\ba}{\begin{eqnarray}}
\newcommand{\ea}{\end{eqnarray}}
\newcommand{\ban}{\begin{eqnarray*}}
\newcommand{\ean}{\end{eqnarray*}}
\newcommand{\al}{\alpha}
\newcommand{\cl}{{\rm class}}
\begin{document}

\maketitle

\parbox{7cm}{\rule{70mm}{1mm} \linebreak
\rule[1mm]{0cm}{0.5cm}
{\sf Relativity Group $\bullet$ Konstanz University \linebreak
\rule[1mm]{0cm}{0.5cm}
$\bigl\langle$ Theory of Gravity $\bigr|$ Quantum Theory
$\bigr\rangle$} \linebreak
\rule{70mm}{1mm}}
\hfill

\vspace{0.5cm}
{\sl Preprint  KONS-RGKU-93-13}

\vspace{1.3cm}

\nopagebreak
\begin{abstract}
The method of restricted path integrals allows one to effectively
consider continuous (prolonged in time) measurements of quantum
systems. Monitoring of the system coordinates is such a
continuous measurement that allows one to describe a quantum system
in terms of trajectories. This approach is applied to chaotic
systems. The behavior of such systems is qualitatively
investigated in classical and quantum regimes of the coordinate
monitoring. The comparison of classical and quantum
chaos in terms of trajectories is performed. Characteristic
features of chaotic systems (observables) with respect to
continuous measurements are analyzed in comparison with those of
regular (non-chaotic) and quantum-nondemolition variables.
\end{abstract}
\section{Introduction}

Characteristic features of chaotic systems (i.e. systems showing
deterministic chaos) are naturally formulated in terms of
trajectories. In the framework of conventional quantum mechanics
the quite different language of wave functions is used. It is usually
claimed that this difference of language is unavoidable
\cite{Zasl}. This is why the problem of quantum
chaos is conventionally formulated as investigation of
characteristic features (for example peculiarities of spectra) of
quantum systems obtained by quantization of chaotic
classical systems 
\cite{Zasl}-\cite{spectra2}. In the present paper
theory of  quantum continuous measurements (in its
restricted-path-integral
version 
\cite{Feynman-path}-\cite{restrict-path-int}) is applied to
investigate
both  classical and quantum properties of the system in terms of
trajectories, i.e. with the help of one and the same language.

One may look at the subject from another point of view and ask
himself how can an experimenter observe deterministic chaos.

Evidently, an experimenter should perform monitoring of some
observable $A$ of the system (during a sufficiently long time) to
obtain the trajectory $[a]=\{ a(t)|0\le t\le T\}$. Then he should
analyze statistical characteristics of this trajectory. It is
important for us that an experimenter begins from performing a
continuous (prolonged in time) measurement of the system.

However it is well known that any measurement of a quantum system
disturbs its state. As a result the evolution of the system
subject to a continuous measurement cannot be described by classical
laws. Why then (and under what conditions) may one talk about
deterministic chaos or chaotic character of classical trajectories?
Have classical trajectories (and specifically chaotic trajectories)
anything to
do with reality? When usage of these trajectories
is correct? What are corrections to the theory of chaos resulting
from quantum features of dynamical systems? Answers to these
questions can be obtained with the help of theory of continuous
quantum measurements because the latter is formulated in terms of
trajectories.

Of course, classical theory and particularly theory of classical
chaos is applicable in a wide class of conditions when the system
may be considered as classical. However this is only an
approximation. Strictly speaking, any system is a quantum one.
It is important to clearly understand when classical approximation
is applicable. Therefore advantageous would be such an approach
which could 1)~supply common language for description of both a
quantum system and its classical approximation and 2)~provide
continuous transition from the conditions when quantum
description is necessary to those when classical approximation is
sufficient.

Theory of continuous quantum measurements

as2,contin
\cite{Feynman-path}-\cite{contin-meas3} is an approach of this type.
This may be seen first of all in the case of such a typical
continuous measurement as monitoring coordinates (position) of
the system. Monitoring of position gives a trajectory as its
output. Therefore a quantum system undergoing position monitoring may
be described in terms of trajectories, the language
characteristic for classical theory.

Continuous transition from the quantum description of the system
and its measurement to a classical approximation (or vice versa)
may be traced when the measurement precision is continuously
changed. In the case of a rough measurement quantum effects are
negligible so that the classical regime of measurement is
realized. When the measurement is precise enough, the quantum
regime takes place with essential quantum effects.

Therefore the program may be naturally formulated to investigate
position monitoring for typical chaotic systems and to analyze the
distribution of the measurement results (formulated in terms of
trajectories) in the case of classical and quantum regimes of
measurement. The first regime should give classical chaos while
the second one may be called quantum chaos. It is essential that
both types of chaos will be described in this case in terms of
trajectories.\footnote{It should be emphasized that the term
AAquantum chaosBB will be used in this paper only in the sense
AAchaos of trajectories (or rather corridors) obtained from
continuous measurement performed in quantum regimeBB.}

One of the most efficient methods for investigating continuous
quantum measurements is a restricted-path-integral method
\cite{Feynman-path}-\cite{restrict-path-int} in which integration
over all
paths in the Feynman integral is replaced by integration over a
restricted set of paths compatible with the measurement output.
We shall apply this method to the problem in question with the
aim to qualitatively analyze the phenomena of quantum and
classical chaos.

We shall see as a result of the analysis that there are systems
showing both classical and quantum chaos in their behavior, the
systems with only quantum chaos and those demonstrating neither
quantum, nor classical chaos.

\section{Quantum Chaos of Trajectories}

The monitoring of position is an (approximate) measurement of
position $q$ in each instant of time. An output of the position
monitoring may be expressed by a trajectory $[a]=\{a(t)|0\le t\le
T\}$. Interpretation of the measurement output is that the
position $q$ in time moment $t$ is close to $a(t)$. Because of a
finite precision $\Delta a$ of the measurement, the coordinate
$q(t)$ may differ from $a(t)$, but not more than by $\Delta a$.
Therefore, adequate representation of the measurement output is
not the trajectory $[a]$ but a corridor $\al$ of the width
$2\Delta a$ centered around $[a]$ (see Fig.~\ref{corridor}).
\begin{figure}
\vspace{6 cm}
\caption{\rm The output of the position monitoring may be denoted by
the trajectory $[a]$ but it is adequately presented by a corridor
having the width $2\Delta a$ equal to the doubled measurement error.
A quantum system undergoing such a measurement should be described by
the path integral with integration restricted on paths lying inside
the corridor.}
\label{corridor}
\end{figure}

The main point of the restricted-path-integral method
\cite{restrict-path-int} is restriction of the Feynman path integral
on
the set of paths lying inside the corridor $\al$. This gives the
probability
amplitude for the given measurement output:
\be
U_{\al}=\int_{\al} e^{\frac{i}{\hbar}S[q]}\,d[q].
\label{path-amplitude-sum}\ee
Monitoring of any observable $A$ may be considered in an
analogous way.

The classical regime of continuous measurement takes place when the
measurement is rough enough (in comparison with some
characteristic quantum threshold which should arise from the detailed
calculation). This means that the corridors
representing measurement outputs are wide enough. In this case
only those measurement outputs have high probability which are
compatible, up to the error of measurement, with the classical
prediction. In the case of the position monitoring only those
corridors $\al$ are probable which contain the classical trajectory
$[a_{\rm class}]$ (see Fig.~\ref{cor-cl-qu}).
\begin{figure}
\vspace{6 cm}
\caption{\rm The measurement is performed in the classical regime if
it is rough (corridors are wide). The regime is quantum if the
measurement is precise enough (in comparison with a certain quantum
threshold). In the classical regime (a) only those corridors may
arise as the measurement outputs which are compatible with classical
predictions, i.e. contain the classical trajectory $[a_{\cl}]$. In
the quantum regime (b) corridors may be incompatible with classical
predictions (demonstrating a quantum measurement noise).}
\label{cor-cl-qu}
\end{figure}

In the quantum regime of measurement, the measurement output may be
incompatible with the classical prediction (may differ from the
latter by more than the measurement error) \cite{restrict-path-int}.
For example, in the case of the quantum regime of position
monitoring, when corridors are narrow, even those corridors that do
not contain the classical trajectory $[a_{\rm class}]$, may arise
with high probability. The more precise is the measurement in this
regime, the further corridors (outputs) may be from the classical
trajectory.

Let us describe this in somewhat more detail taking monitoring
of position (or of another observable) as an example of a continuous
measurement. In this case the curve $[a]$
characterizing the measurement output, may differ from the
classical trajectory $[a_{\rm class}]$ by some value $\delta a$.
In the classical regime of measurement $\delta a$ is equal
to the measurement error $\Delta a$ but it may be much more in the
quantum regime. Moreover, in the quantum regime a paradoxical
situation arises: the less is the measurement error $\Delta a$,
the more is the variance of the measurement outputs $\delta a$. This
is a typical consequence of unavoidable back reaction of the
measuring device onto the measured quantum system.

This deflection of the measurement outputs from classical
predictions is nothing else than the quantum measurement noise.
In the present context one may call this phenomenon by {\em quantum
chaos}.

\section{Regular and Chaotic Observables}

Now we can compare, from the point of view of
continuous measurements, regular systems with chaotic ones.

It follows from the argument of the preceding section that,
in the classical regime of measurement, the outputs coincide
(up to the measurement error) with those predicted by
classical theory. Therefore, if the system is regular (in the
sense that deterministic chaos is absent for such a system), it
behaves regular when being observed in the classical regime. Such a
system has no classical chaos.\footnote{\label{reg-chaot-obs}This
affirmation
concerns also chaotic systems if a specially chosen regular
observable is measured.} This is of course almost
tautology. In the quantum regime of measurement
such systems behave chaotic because of the quantum measurement
noise (see the preceding section).

Consider now chaotic systems i.e. those that have chaotic
classical trajectories. It is evident that such systems behave
chaotic in the classical regime of measurement (by definition
of a chaotic system). What then may be said about the quantum
regime?

We saw in  the preceding section that, in the quantum regime,
even those corridors possess high probability which are far from
classical trajectories. Hence the direct connection of the
measurement outputs with classical trajectories is absent. The
question arises whether it is possible that no quantum chaos exists
for such systems. It would take place if the high probability
corridors could form some regular families leading to no chaos
in laws governing these families.

However in reality this is not the case. The reason is that each
corridor containing a classical trajectory, has high
probability. In other words, a classical trajectory being inside
the corridor is not (in the quantum regime) necessary but it is
sufficient condition for the corridor having high probability.
The reason is in the fact that the action functional has its
extremum on a classical trajectory.

Indeed, the probability is comparatively low when the action
$S[q]$ in the Feynman exponentials
$$
\exp\left (\frac{i}{\hbar}S[q]\right )
$$
changes quickly for $[q]\in \al$ so that destructive interference
arises. If the variation of the action is slow for paths belonging to
the given corridor $\al$, then the exponentials  sum up to give
an amplitude of comparatively large absolute value. This is valid
even for a corridor that contains a subset (having sufficient
measure) of paths with slowly varying action. The situation is
just this if the corridor has a classical trajectory inside
it.\footnote{Of course, this statement should be formulated more
correctly from the mathematical point of view and proved strictly.
This is a very interesting and not easy task. However it is clear on
physical ground that it is valid in typical situations.}

One sees from this argument that each classical trajectory
determines a corridor having comparatively high probability. the set
of all high-probability corridors turns out to be richer than the set
of classical trajectories. But classical trajectories are chaotic in
the considered case of classically chaotic systems.
Therefore chaos of classical trajectories results in chaos of
corridors, i.e. chaos of the measurement outputs.

The conclusion that may be extracted from this argument is that a
classically chaotic system (or rather a chaotic variable of such
a system) possesses also quantum chaos (i.e. the chaotic character
of outputs of a continuous measurement performed in the quantum
regime).

Summing up, we see that there are systems (observables)
possessing only quantum chaos (regular systems or observables)
and those possessing both classical and quantum chaos (chaotic
systems or observables). It may be thought that quantum chaos is a
common feature of all systems so that there exist no observable
without quantum chaos. This however is not the case. We shall see
in the next section that there is a class of observables
(so-called quantum nondemolition, QND, observables) that possess
no quantum regime of measurement and therefore no quantum chaos.

\section{Quantum Nondemolition Measurements}

Let us consider the physical reason for quantum measurement
noise. It can be formulated as disturbing a canonically conjugate
observable.

Quantum measurement noise, i.e. deflection of corridors
(measurement outputs) from the classical trajectory, may be
formulated as disturbing evolution of the system because of the
measurement. More precisely, evolution (time dependence) of the
measured observable is disturbed when it is measured  in
quantum regime. Why does this occur? The reason is that
the measurement of an observable, because of the uncertainty
principle, disturbs its canonically conjugate observable.

This may be illustrated by a simple consideration. If one
measures (with a finite precision) the coordinate $q$ of a free
particle in some instant, this measurement disturbs the linear
moment $p$. This in turn disturbs further evolution of $q$
because of the equation of motion $m\dot q=p$. The same is
usually valid for any pair of canonically conjugate observables.

However there are observables of a special nature (so-called quantum
nondemolition, QND, observables) such that measuring them does
not influence their evolution 
\cite{QND1}-\cite{Brag-book}. What does occur in this case? When one
measures a QND  observable $X$, its canonically
conjugate observable $Y$ is unavoidably disturbed as a result of the
uncertainty principle. However a QND observable differs from a
generic one in that its dynamics does not depend on the value of the
canonically conjugate observable. For example, the following equation
may be fulfilled for a QND variable:
\be
\dot X = f(X).
\label{QND-evolut}\ee
The right-hand side of this equation does not depend of the
observable $Y$, canonically conjugate to $X$. Therefore disturbance
of $Y$, resulting from the measurement of $X$, does not change
evolution of $X$.

A linear momentum $p$ of a free particle is an example of a QND
observable, since it satisfies the equation $\dot p = 0$. The
momentum $p$ is a QND variable even for a particle under action of
external force $F(t)$ not depending on the particleBs position $q$
(because $\dot p=F/m$ in this case). One more, and less trivial
example is a pair of (canonically conjugate to each other) quadrature
components of a harmonic oscillator:
\ban
X&=&q\cos\omega t -\frac{p}{m\omega}\sin\omega t,\\
Y&=&q\sin\omega t +\frac{p}{m\omega}\cos\omega t.
\ean
Each of them is a QND variable.

The consequence of such a feature of QND variables is that there is
no quantum regime (and therefore no quantum chaos) in their
monitoring, even if an arbitrarily precise measurement is performed
during the monitoring \cite{QND-path-theory}. Being regular
(non-chaotic), QND variables have no classical chaos too. Therefore,
observables of this class possesses neither classical, nor quantum
chaos. One may doubt that QND variables are necessary regular.
However this may be proved in the following way.

Let $X$ is a QND observable in a system having in general chaotic
variables. May $X$ be also chaotic or not? Being QND, the variable
$X$ has specific features. The characteristic feature of the dynamics
of such a variable is that the function of time $X(t)$ is
unambiguously determined by $X(0)$. This means that the equation
(\ref{QND-evolut}) is valid. Therefore there is one-dimensional
subsystem (with the coordinate $X$), in the system under
investigation, having completely autonomous dynamics. Being
one-dimensional, this subsystem cannot be chaotic, even if it is
non-linear. Therefore a QND variable is necessarily regular.

Let us say several words about terminology. We discussed in detail
behavior of QND observables and showed that they possess very special
properties demonstrating no chaos at all. Ordinary observables (such
as the coordinate of a regular system) show, as it has been discussed
in the section ``Quantum Chaos of Trajectories'', no classical
chaos, but they show quantum chaos. The variables of this class
may be called (to distinguish them from QND ones) quantum
demolition (QD) observables.

Chaotic systems or rather chaotic observables (i.e. those showing
chaotic behavior when considered classically) are in a sense
opposite to the case of QND variables. Evolution of an observable
of a chaotic system depends exponentially on the values of this
observable and of its conjugate.\footnote{This concerns generic
observables of chaotic systems but this is not valid for specific
regular observables existing in them (see footnote
\ref{reg-chaot-obs} above).} Therefore the measurement (performed in
quantum regime) of such an observable must disturb its evolution more
strongly than in the case of  regular QD observables.. Observables
(measurements) of chaotic systems may be called strongly quantum
demolition (SQD) ones.

The chain is thus naturally determined, of QND, QD, and SQD (or
chaotic) variables. QND variables are completely regular, i.e. they
show no chaos in any regime of measurement. In fact no quantum regime
of measurement exists for such observables. QD (usual) variables are
classically regular but chaotic in the quantum regime of measurement.
At last, SQD (chaotic) variables are chaotic both in quantum and
classical  regimes of measurement. This is illustrated by
Table~\ref{chaos-table}.
\begin{table}
\caption{Different types of observables show regular or chaotic
behavior in classical and quantum regimes of
observation.\label{chaos-table}}
\begin{center}
\begin{tabular}{|c|c|c|}\hline
	&classical regime	&quantum regime\\\hline
QND	&regular	&regular\\
QD	&regular	&chaotic\\
SQD	&chaotic	&chaotic\\
\hline
\end{tabular}
\end{center}
\end{table}

It is worthwhile to make one more remark. Though we talked for
simplicity about chaotic observables (or even chaotic systems), in
reality these observables have usually areas of regularity. In the
limits of such an area the observable have all properties of
a regular one. It may turn out to be ordinary (QD) observable or
even QND observable while its measurement gives outputs in the
regular area.

\section{Conclusion}

We have analyzed in this paper the behavior of chaotic systems
from the point of view of continuous measurements with quantum
effects taken into account. In most arguments monitoring of some
observable $A$ was taken as an example of continuous measurements. An
output of such a measurement is presented by a trajectory $[a]=\{
a(t)|0\le t\le T\}$ or rather by a corridor of paths centered around
the trajectory. (The quantum system undergoing the measurement is
described by the Feynman path integral with integration over the
corridor).

Thus even a quantum system is described by trajectories, and chaos
may be described by statistics of trajectories both for quantum as
well as for classical systems.

Two regimes of the measurement (classical and quantum regimes) were
considered and three types of observables were distinguished as a
result of the analysis: quantum nondemolition (QND), ordinary (or
quantum demolition, QD), and chaotic (or strongly quantum demolition,
SQD) observables.

The main goal of the paper was to found out whether chaotic behavior
takes
place in both considered regimes of measurement. In other words, the
question was whether the system shows classical chaos and/or quantum
chaos (chaos of trajectories or corridors is meant in both cases).

The answer turned out to be different for the mentioned three types
of observables: QND variables have neither quantum, nor classical
chaos, QD variables have only quantum chaos, and SQD variables have
both classical and quantum chaos (in all these formulations quantum
chaos of trajectories is meant).

It should be stressed that the analysis presented here is only
preliminary and purely qualitative. Of course, much more detailed
investigation is necessary. Because on nonlinearity of chaotic
systems such an investigation will require numerical techniques or
simulations  (see \cite{OnofrioPRL} on the methods of such
calculations).

\vskip 5mm\centerline{ACKNOWLEDGEMENT}

The author is indebted to Prof.J.Audretsch for his kind hospitality
in the University of Konstanz where this paper was finished. The work
was supported in part by the Deutsche Forschungsgemeinschaft.
Discussions with Dr.R.Onofrio were useful for improving some unclear
parts of the paper.

\null
\end{document}